\documentclass[prb,aps,twocolumn,floats]{revtex4}
\usepackage{amstext}
\usepackage{amsmath}
\usepackage{graphics}
\usepackage{multicol}
\usepackage{epsfig}

\def\identity{{1}}

\def\bfC{{\bf C}}
\def\bfY{{\bf Y}}
\def\a12{\mbox{$\alpha_{12}$}-B}
\def\aortho{\mbox{$\alpha_{ga}$-B} }
\def\bortho{\mbox{$\beta_{ga}$-B} }
\def\ind4{\begin{smallmatrix}\mbox{{\tiny at,pr}} \\ \mbox{{\tiny at',pr'}}\end{smallmatrix}}

\begin{document}

\title{New {\em ab initio} approach for high pressure systems with application to a new high-pressure phase for boron: perturbative momentum-space potentials.}

\author{D.E. Segall\footnote{ Current Address: Department of Applied
Physics, California Institute of Technology, Pasadena, California
91125.}} \address{Department of Physics, Massachusetts Institute of
Technology, Cambridge MA 02139}

\author{T.A. Arias}  \address{Laboratory of Atomic and Solid State Physics, Cornell University,
Ithaca, NY 14853}

\begin{abstract}
Through the use of perturbation theory, in this work we develop a
method which allows for a substantial reduction in the size of the
plane-wave basis used in density-functional calculations. This method
may be used for both pseudopotentials and all-electron calculations
and is particularly beneficial in the latter case.  In all cases, the
approach has the advantage of allowing accurate predictions of
transferability errors for any environment. Finally, this method can
be easily implemented into conjugate gradient techniques and it is
therefore computationally efficient.  In this work, we apply this
method to study high pressure phases of boron. We find that boron
undergoes a phase transition from the \a12 structure to the \aortho
structure, both of which are semiconducting. The \aortho structure has
lower energy than traditional mono-atomic structures, which supports
the assertion that the metallic, and hence superconducting phase, for
boron is much more complicated than a simple mono-atomic crystal.
\end{abstract}

\twocolumn

\maketitle

\section{Introduction} \label{sec-intro}

Experimental techniques are now able to probe condensed matter systems
at higher and higher pressures through the use of diamond-anvil cells
or dynamical shock methods~\cite{exp_rev,eremets}. At these newly
attainable pressures, structural and electronic phase transitions can
occur, opening the door to exploration for new physical phenomena.

Such systems offer an exciting avenue for first principle
calculations, as their predictions can not only follow but also
sometimes lead results from new experimental
techniques~\cite{neaton1,neaton2,struzhkin,fortov,hanfland,vohra,mailhiot,eremets,zhao}.
When applying traditional first principle calculations to such
systems, care must be taken, as most basis sets take advantage of the
distinction between the core regions and valence regions which is
prominent at ambient conditions. However, when studying systems over a
wide range of pressures, this distinction vanishes, and both regions
need to be treated on an equal footing.  The following section
discusses some problems that can occur when applying some of the most
popular basis sets to high pressure systems.

To overcome the above potential pitfalls and to allow for accurate
description of both the core and valence regions while keeping the
mathematical stability and systematic convergence of the plane-wave
basis, we have developed a method based on perturbation theory. This
method can be used for direct all-electron calculations, dramatically
reduces the size of the plane-wave basis while properly accounting for
all higher momentum states, accounts accurately for the core electrons
and their interaction with the valence region, and automatically
generates pseudopotentials which change with the environment thereby
enhancing transferability.  Moreover, the effect of pseudizing in the
crystal can be quantified; hence, accurate predictions for
transferability can be made in {\em any} environment.  Finally, the
method can be implemented into current state of the art minimization
techniques, such as the conjugate gradient method, and is therefore
computationally efficient.

As an application, we study high pressure phases of boron. Boron
crystallizes into many complex structures, which are governed by the
regular icosahedron~\cite{donohue,brs}. Boron is semiconducting at
ambient conditions, turns metallic under pressure ($\approx$ 170 GPa)
and superconducting above 160 GPa~\cite{eremets}. Theoretically,
Mailhiot and coworkers~\cite{mailhiot} were the first to apply
density-functional techniques to the study of the phases of boron at
high pressures.  To date, however, such studies~\cite{mailhiot,zhao}
have been limited to a modest selection of phases and were based on
traditional techniques which artificially separate the physics in the
core and valence regions.  In this work, we apply our new, unbiased
technique to a wider range of phases and find a new phase, the \aortho
structure, to be lower in energy at high pressure than any phase
reported previously and to have important potential implications for
the observed semiconductor-metallic phase transition associated with
the superconductivity.

We proceed as follows: Section~\ref{sec-methods} describes the various
methods that can be used to study systems under high pressure,
specifically, the pros and cons of each method. Then in
Sections~\ref{sec-pert}~and~\ref{sec-cg} our method is developed.
Computational details are presented in Section~\ref{sec-details},
Finally, we apply our approach to study the high pressure phases of
boron in Section~\ref{sec-boron}.

\section{Methods} \label{sec-methods}

Within current state of the art electronic structure calculations,
various approaches can be taken, mainly differentiated by the choice
in basis set. Each set has its own benefits and disadvantages. Some
of the most common basis sets are the plane-waves~\cite{cohen},
Gaussian~\cite{gaussian}, linearized methods~\cite{singh,lmto}, and
wavelets~\cite{arias_wl}. We will now go over the various basis sets
and focus on problems that may occur when each is applied to high
pressure systems.

The Gaussian basis set expands the wave functions in terms of linear
combinations of Gaussians designed to represent the atomic orbitals
well. Such a basis set has the advantageous property of expanding the
wave functions in a small number of computationally efficient basis
functions.  There are, however, some drawbacks. Notably, such a basis
set cannot be improved systematically. This does not tend to be a
practical problem for normal systems; however, such a problem can
manifest itself for a system studied at various volumes, as the basis
set will span a different percentage of real space for systems whose
volume differ appreciably. Because it is unclear of how to convergence
such a basis set systematically, it is unknown whether standard
Gaussian basis sets will perform well at high-pressures, particularly
because such bases are biased toward orbitals constructed at ambient
conditions. It is thus unclear if such a bias can hamper calculations
at high pressures, where unknown phenomena may occur.

Linearized methods generally fall under one of two methods: the Linear
Muffin Tin Orbital (LMTO)~\cite{lmto} method or the Linearized
Augmented Plane-Wave (LAPW)~\cite{singh} method. Of these two, the
LAPW is more accurate and we will therefore concentrate on this
approach.  When studying systems under high pressure, the most notable
problem that can occur with the LAPW method is the treatment of the
core electrons. At high pressures, certain core electrons need to be
promoted and treated as valence electrons, and care must be taken with
this method when promoting such electrons. Two widely used approaches
to promote such electrons are the multiple window
approach~\cite{singh} and the localized orbital
approach~\cite{singh}. Both of these approaches could be particularly
problematic for high pressure systems, particularly for first row
elements. The multiple window method is not guaranteed to give a good
solution~\cite{singh}, as different basis sets are used to expand the
core and valence electrons. In the localized orbital approach, this
problem is resolved by using one energy window. However, problems may
exist, as there are very large energy separations between the valence
bands and the core bands, particularly for first row elements.

The plane-wave approach, when used with the pure Coulomb potential,
has the advantageous property that convergence of the energy can be
systematically improved in a stable and controllable
framework. Moreover, such a basis is unbiased and can therefore
represent core and valence regions on an equal footing.  However, such
an approach has only been used for the simplest systems, such as
hydrogen~\cite{barbee,natoli,ballaiche} and lithium~\cite{ballaiche}
as it requires a huge number of plane-waves to properly account for
the singular Coulomb potential and the exclusion principle. Therefore
such an approach is not practical beyond the simplest systems.

The pseudopotential approach had been developed to circumvent the
above difficulties with plane waves
bases~\cite{hamann,kleinman,pickett}.  However, pseudopotentials are
poorly suited for studying high pressure systems because, although
some key properties are known~\cite{hamann,teter,rappe}, the
transferability of pseudopotentials is not systematically understood,
particularly when the pseudized regions occupy a large percentage of
the available volume as they do at high densities.  More
fundamentally, apart from nonlinear core corrections in the
exchange-correlation potential~\cite{louie}, pseudopotentials
completely ignore the core region, which have been shown to play a
vital role for high pressure systems~\cite{neaton1,neaton2}.

One of the most successful techniques, which keeps the benefits of the
pseudopotential approach and can allow for core electrons and can
recover details of the true wave function in the core region, is the
Projected Augmented Wave (PAW)~\cite{paw} method. This method has been
used to study Lithium and Sodium~\cite{neaton1,neaton2} under high
pressure.  Such a method is quite elegant; however, it must be
constructed from a referenced atomic state. Moreover, the construction
of such a potential still requires a real space cutoff. In order to
effectively study high pressure systems, such a cutoff should be made
quite small, so that the pseudized regions do not overlap.  The
approach then looses many of its computational benefits.

The final type of basis set of which we are aware in the study of
solid state systems are wavelet bases~\cite{arias_wl}. Such bases have
the attractive property of using multi-resolution analysis to provide
the necessary resolution in each region of space in a systematic and
mathematically stable manner and thereby efficiently handle valence
and core electrons exactly.  However, such a basis set has just come
to the fore and has not yet been adopted by many groups.

We now describe a technique which has the advantage of using a
plane-wave basis without pseudopotentials. This method has the added
benefit of greatly reducing the size of the plane-wave basis set by
generating the important high momentum states through perturbation
theory.  Below, we show that a properly constructed perturbation
expansion can be quite accurate in this context.  Finally, we show
that this method can easily be implemented in current conjugate
gradient techniques and therefore can be used with highly optimized
minimization techniques.

\section{Perturbative Derivation}\label{sec-pert}

In the next two sections the method will be derived.  To gain insight
into this method, it will first be derived for the simple case of
directly diagonalizing the single-particle Hamiltonian. Then,
connections will be made to previous approach~\cite{lowdin}. In
Section~\ref{sec-cg}, the procedure will be implemented into the
conjugate gradient framework, so that optimized calculations can be
performed. Technical details will be discussed in
Section~\ref{sec-details}.  For simplicity in the formal developments
below, we sample the Brillouin zone at the Gamma point.  Adding in
k-point dependence is straightforward.

\subsection{Derivation for fixed potentials}~\label{sec-diag}

Standard electronic structure calculations seek for the minimum of an
energy functional $E[\{C\}]$ as a function of a set of basis function
coefficients $\{C\}$ subject to an orthonormality constraint.  The
variational derivative of this energy functional and constraint then
lead to the standard eigenvalue problem
\begin{equation}
H C = \epsilon C, \label{eqn:sd_ham}
\end{equation}
where $H$ is the single-particle Hamiltonian (fixed and not updated
self-consistently for the discussion in this section), $C$ is an
eigenvector and $\epsilon$ is the corresponding eigenvalue.  Note that
we here assume that the basis set is orthonormal as our objective here
is to work with plane waves.

Separating the coefficients into two groups, those describing behavior
in the low spatial frequency space $P$ as $C^{P}$ and those describing
behavior in the high spatial frequency space $Q$ as $C^{Q}$, also
separates the couplings which $H$ describes into four groups,
couplings from low-frequency to low-frequency $H^{PP}$, from low- to
high- frequency $H^{QP}$, from high to low $H^{PQ}$ and from high to
high $H^{QQ}$.  The above equation then decomposes simply into
\begin{eqnarray}
H^{PP} C^{P} + H^{PQ} C^{Q} & = & \epsilon C^P \label{eqn:P_ham} \\
H^{QP} C^{P} + H^{QQ} C^{Q} & = & \epsilon C^Q. \label{eqn:Q_ham}
\end{eqnarray}
Solving for the $Q$-space components of the
eigenfunction in terms of the $P$-space components leads to
\begin{equation}
C^Q = - \frac{1}{H^{QQ} - \epsilon}H^{QP} C^P, \label{eqn:CQfull}
\end{equation}
where the fraction means the inverse of an operator.

A standard technique~\cite{feshbach,lowdin,atomic} is to substitute
Equation~(\ref{eqn:CQfull}) into Equation~(\ref{eqn:P_ham}) in order
to generate an effective eigenvalue equation for the $P$-space,
\begin{equation}
(H^{PP} + H^{PQ}\frac{1}{\epsilon - H^{QQ}}H^{QP})C^P = \epsilon C^P.
\label{eqn:oldway}
\end{equation}
This equation reduces the problem to the $P$-space only. If standard
diagonalization techniques are used, this reduces the time to
diagonalize the Hamiltonian by a factor of $(\frac{N^P}{N^P+N^Q})^3$,
where $N^{\alpha}$ is the number of basis functions in the
$\alpha$-space. However, it is well known that such a decomposition
does not necessarily simplify the problem at hand, particularly
because inversion of a matrix is computationally intensive and care
must be taken in order to achieve self-consistency in the eigenvalue.
We now describe an efficient way to go beyond the standard approach to
address each of these issues in turn.

In a plane-wave basis, the inversion of $\epsilon - H^{QQ}$ can be
approximated accurately and simply in a way which will save
considerable computational effort, especially when used in conjunction
with conjugate gradient techniques.  Specifically, because the
$Q$-space contains the high momentum plane-wave states, the kinetic
energy dominates $H^{QQ}$, which we may then approximate to be the
kinetic energy operator $H^{QQ}_o$, whose $\alpha,\beta$ component is
\begin{equation}
H^{QQ}_{o,\alpha \beta} = \frac{1}{2}\vec Q^2_{\alpha}
\delta_{\alpha,\beta}. \label{eqn:hqq}
\end{equation}
Here $\delta_{\alpha,\beta} $ is the Kronecker delta and $\vec Q$ is a
reciprocal lattice vector, and atomic units have been assumed.

In principle, the eigenvalues appearing in the left side of
Equation~(\ref{eqn:oldway}) must be calculated self-consistently, but
this leads to difficulties.  Each eigenstate sought then sees a
different Hamiltonian and orthonormality is lost.  Alternately, in
practice one employs on the left-hand side a constant, approximate
$\epsilon$ which one hopes to be appropriate for all desired
eigenvalues.  For us, neither approach is satisfactory.  To circumvent
these difficulties, we linearize $C^Q$ in terms of the eigenvalue
\begin{equation}
C^Q = -\frac{1}{H^{QQ}_o}(\identity + \frac{\epsilon}{H^{QQ}_o})H^{QP}
  C^P + \mathcal{O}(\frac{\epsilon}{H^{QQ}_o})^2. \label{eqn:CQtemp}
\end{equation}
Note that in a plane-wave basis such an expansion will converge
quickly, as the cutoff in the $P$-space is, in general, much larger
than the eigenvalues of interest.

Substituting Equation~(\ref{eqn:CQtemp}), to linear order in
$\epsilon$, into Equation~(\ref{eqn:P_ham}), gives an effective
generalized eigenvalue equation,
\begin{equation}
(H^{PP} + H^{eff}) C^P = \epsilon \mathcal{O} C^P. \label{eqn:ham_eff}
\end{equation}
Here
\begin{equation}
H^{eff} = - H^{PQ} \frac{1}{H^{QQ}_o} H^{QP} \label{eqn:Heff}
\end{equation}
is the effective single-particle Hamiltonian and
\begin{equation}
\mathcal{O} = \identity + H^{PQ} \frac{1}{H^{QQ}_o}\frac{1}{H^{QQ}_o} H^{QP}
\label{eqn:Overlap}
\end{equation}
is a positive-definite overlap matrix.  This equation will generate
errors of order $(\epsilon/E^P_c)^2$ in the eigenvalue, where $E^P_c$
is the cutoff energy for the $P$-space.  We below, show that this
truncation error proves to be a good measure of the error in the
energy per atom.  Thus, our method provides for accurate {\em a
priori} predictions for transferability errors.

In addition to computing the eigenvalue spectrum, we often require
access to the eigenstates, particularly for density functional theory
calculations which require self-consistent solution of the electronic
states within a potential dependent upon those states.  Calculation of
the self-consistent potential requires not only the $P$-space
components $C^P$ from (\ref{eqn:ham_eff}) but also the $Q$-space
components.  The most direct choice for generating the $C^Q$ is to use
Equation~(\ref{eqn:CQtemp}); however, we wish to eventually apply
this technique to the conjugate gradient method and defining $C^Q$ as
in Equation~(\ref{eqn:CQtemp}), no longer preserves orthonormality
for the full eigenvector set $\{C\}$ and post-reorthonormalization can
make the conjugate gradient procedure unstable.

To avoid these difficulties, we note that the solutions to
the generalize eigenvalue problem (\ref{eqn:ham_eff}) automatically satisfy
\begin{eqnarray*}
\delta_{ij} & = & C^{P\dagger}_i \mathcal{O} C^P_j \\
& = & C^{P\dagger}_i \left( \identity + H^{PQ} \frac{1}{H^{QQ}_o}\frac{1}{H^{QQ}_o}
H^{QP} \right) C^P_j
\end{eqnarray*}
where the subscripts indicate the coefficients for individual states
$i$ and $j$ and we have substituted the definition
(\ref{eqn:Overlap}).  Regrouping terms, we find that identically
\begin{eqnarray*}
\delta_{ij}
& = & C^{P\dagger}_i C^P_j + \left(-\frac{1}{H^{QQ}_o} H^{PQ}
C^P_j\right)^\dagger \left(-\frac{1}{H^{QQ}_o} H^{PQ}
C^P_j\right),
\end{eqnarray*}
so that the set of complete states $\{C\}$ will be exactly orthonormal
provided we make the identification,
\begin{equation}
C^Q \equiv -\frac{1}{H^{QQ}_o}H^{QP} C^P \label{eqn:CQ},
\end{equation}
which is nothing other than (\ref{eqn:CQtemp}) truncated at zeroth
order.  We thus conclude that to avoid issues of orthonormality to
allow simple implementation of conjugate gradient techniques, one {\em
must} construct $C^Q$ at this order.  Accordingly, we employ
(\ref{eqn:CQ}) as the construction for $C^Q$ throughout the reminder
of this work.

\subsection{Connections with L\"{o}wdin Perturbation Theory}

To carry out full density-functional calculations based upon the
results from the previous section, one in principle first would solve
(\ref{eqn:ham_eff}) using direct diagonalization techniques to find
$C^P$ and then construct $C^Q$ from (\ref{eqn:CQ}).  Next, from the
wave functions, one would update the charge density, Hartree and
exchange-correlation potentials, recompute $H^{PP}, H^{eff}$ and
$\mathcal{O}$.  Finally, one would iterate this procedure to
self-consistency.  As this approach is quite similar to the L\"{o}wdin
perturbation theory~\cite{lowdin} which was used in the mid 1980's by
a number of groups~\cite{joanop_ld,cohen_ld} with regard to electronic
structure calculations, this section briefly reviews the L\"{o}wdin
approach as described in Reference~\onlinecite{joanop_ld} and discuss the
major differences between that approach and ours.  The following
section, Section~\ref{sec-cg}, shows how conjugate gradient
techniques may be applied directly to our approach but not to
L\"{o}wdin perturbation theory.

L\"{o}wdin perturbation theory also decomposes behavior into high and
low momentum plane-wave states.  This approach also solves an
eigenvalue problem 
$$
U^{PP}C^P = \epsilon C^P
$$ 
for a renormalized Hamiltonian $U^{PP}$ similar to that of
Equation~(\ref{eqn:oldway}),
\begin{equation}
U^{PP} = H^{PP} +  H^{PQ} \frac{1}{\epsilon - H^{QQ}_o} H^{QP},
\label{eqn:low_ham}
\end{equation}
Self-consistency is reached by calculating the charge density either
with $C^P$ alone, or with both $C^P$ and $C^Q$, with
\begin{equation}
C^Q = \frac{1}{\epsilon - H^{QQ}_o}U^{QP}C^P. \label{eqn:low_CQ}
\end{equation}
Here $U^{QP}$ is similar to $U^{PP}$, except that the first index is
in the $Q$-space.  For the reasons discussed above in
Section~\ref{sec-diag}, if $C^Q$ is employed, then the wave functions must
be re-orthonormalized.

In practice\cite{joanop_ld}, the value employed for $\epsilon$ in
Equation~(\ref{eqn:low_ham}) depends upon the physics under
exploration.  Thus, prior to the calculation, the shape of the band
structure should be approximately known and should have a relatively
narrow band-width, requiring great care for the study of systems under
unusual physics conditions such as extremely high pressures.
Moreover, the approximation of $\epsilon$ by a constant $\tilde
\epsilon$ is always a first order error in the solution to the
eigenvalue. This can simply be seen by Taylor expanding
Equations~(\ref{eqn:low_ham})~and~(\ref{eqn:low_CQ}) in a similar
fashion as Equations~(\ref{eqn:CQtemp})~and~(\ref{eqn:ham_eff}). If
$\tilde \epsilon$ is close to $\epsilon$, then this error is small as
is the next order. However, if there is a large discrepancy between
$\epsilon$ and $\tilde \epsilon$, then these errors are {\em
manifestly} first order.  This is particularly troublesome for
calculations including core electrons where the core states energies
have large separations from the valence states.  We thus expect
L\"{o}wdin perturbation to be useful primarily only within a
pseudopotential framework.  In our approach, by contrast, {\em all}
band energies are treating on an equal footing to second order.  This
is because Equation~(\ref{eqn:ham_eff}) includes all terms through
first order by shifting some terms to the right-hand side to form the
generalized eigenvalue problem.

The great advantage of L\"{o}wdin perturbation theory for its time is
that it reduces the time for standard diagonalization from
$(N^P+N^Q)^3$ to $(N^P)^3$.  However, in the mid 1980's standard
diagonalization techniques were dropped in lieu of much more efficient
Car-Parrinello~\cite{cp,payne} and conjugate
gradient~\cite{teter_cg,payne,arias_cg} techniques, which, being based
on minimization, require the eigenvalue equations to represent the
variational derivative of some energy functional.  It is not clear
that the Hamiltonian used within L\"{o}wdin perturbation theory is a
good approximation to the variational derivative of an energy
functional. What particularly complicates this is the dependence of
the Hamiltonian on the eigenvalue and that the unfolding of the wave
vector into the $Q$-space does not preserve orthonormality. Therefore,
it is unclear how one would employ L\"{o}wdin perturbation theory in
conjunction with such optimized minimization techniques.  The next
section demonstrates that, in contrast, our approach represents a very
good approximation to the variational derivative of an energy
functional and, because it preserves orthonormality, can be used
directly with optimized minimization techniques.

\section{Applications To  Conjugate Gradient Techniques}\label{sec-cg}

We first discuss applicability to a traditional conjugate gradient
techniques in Section~\ref{sec:trad} and generalize to the analytic
continued approach~\cite{arias_cg,dftpp} in
Section~\ref{sec:analcont}.  Section~\ref{sec-details} gives details
of our computational implementation and provides specific
computational studies of the various approximations and the
convergence of the approach.

\subsection{Application to Traditional Conjugate Gradient Techniques}
\label{sec:trad}

Again, we seek the minimum of an energy functional of the form
$E[\{C\}]$ as a function of a set of wave function coefficients
$\{C\}$.  To formulate the separation into low- and high- spatial
frequency components as a variational principle associated with an
energy functional, we express the energy functional directly in terms
of the $\{C^P\}$ and $\{C^Q\}$ {\em before} taking any variations,
$E[\{C^P\},\{C^Q\}]$.  We now seek the minimum of this functional
subject to the orthonormalization condition
\begin{equation}
C^{P\dagger}_i \mathcal{O} C^P_j = \delta_{ij}. \label{eqn:oncond}
\end{equation}
Here $\mathcal{O}$ is an overlap matrix, which should give the correct
overlap matrix at the minimum of the energy
functional. Section~\ref{sec-details} will describe the overlap matrix
that we use in our calculations. Finally, the $\{C^Q\}$ are left
{\em unconstrained}.

The sets $\{C^P\}$ and $\{C^Q\}$ are treated as independent variables.
However, at each $\{C^P\}$ point, the functional is directly minimized
with respect to the set $\{C^Q\}$. The gradients are
\begin{eqnarray}
\frac{\partial E}{\partial C^{P\dagger}_i} & = & H^{PP} C^P_i + H^{PQ}
C^Q_i - \epsilon \mathcal{O} C^P_i \label{eqn:gradPold}
\end{eqnarray}
and
\begin{eqnarray}
\frac{\partial E}{\partial C^{Q\dagger}_i} & = 0 = & H^{QP} C^P_i +
H^{QQ}C^Q_i, \label{eqn:gradQ}
\end{eqnarray}
where the second term is set to zero, emphasizing that the energy
functional is minimized with respect to $\{C^Q\}$ at each point
$\{C^P\}$.

\begin{figure}
\begin{center}
\scalebox{0.4}{\includegraphics{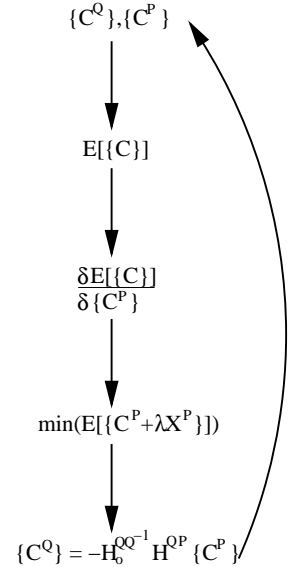}} \\
\end{center}
\caption{Flow chart for application to traditional conjugate gradient
techniques.}
\label{fig:flowchart}
\end{figure}

The flowchart in Figure~\ref{fig:flowchart} describes the minimization
procedure within the traditional conjugate gradient framework.  Given a
point at $\{C^P\}$ and $\{C^Q\}$, calculate the energy. Next, calculate
the gradient for the set $\{C^P\}$,
Equation~(\ref{eqn:gradPold})~or~(\ref{eqn:gradP}), and hold the set
$\{C^Q\}$ fixed.  Minimize the energy functional for the set
$\{C^P\}$ only, along the conjugate direction $\{X^P\}$.  After the
functional is minimized along this particular direction, the
functional is then minimized with respect to $\{C^Q\}$,
Equation~(\ref{eqn:gradQ}), by setting
\begin{equation}
C^Q = -\frac{1}{H^{QQ}_o}H^{QP} C^P. \label{eqn:CQmin}
\end{equation}
Finally, this process iterates until the energy functional reaches its
minimum.

In calculating ${C^Q}$, we have again approximated $H^{QQ}$ by
$H^{QQ}_o$, and therefore do not exactly minimize the functional with
respect to $C^Q$.  Although the form for $C^Q$ is not that which
exactly minimizes our functional, we have found the approximation to
be sufficiently close that the conjugate gradient technique is quite
stable and efficient.  Using such a form for $C^Q$, the gradient for
$C^P$ indeed becomes
\begin{eqnarray}
\frac{\partial E}{\partial C^{P\dagger}_i} & = & H^{PP} C^P_i + H^{eff}
C^P_i - \epsilon \mathcal{O} C^P_i, \label{eqn:gradP}
\end{eqnarray}
where $H^{eff}$ is as defined in Equation~(\ref{eqn:Heff}).  Thus, we
find that, unlike the L\"{o}wdin perturbation theory approach, the
prescription in Section~\ref{sec-diag} represents a set of equations
that is, to a very good approximation, a variational principle and
therefore to be amenable to solution with conjugate gradient methods.

\begin{figure}
\begin{center}
\mbox{\rotatebox{90}{ {\hspace*{.3in}
      $log\left(\frac{(E-E_{min})}{\left| E_{min} \right|} \right)$}}
\scalebox{0.3}{\includegraphics{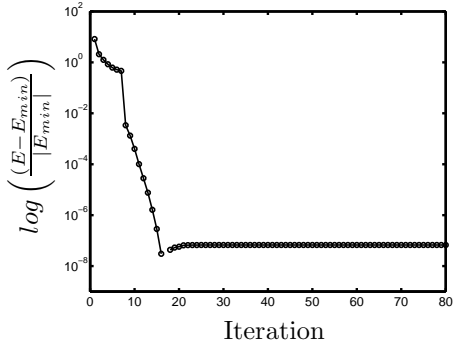}}} \\
{ \hspace*{.4in} Iteration}
\end{center}
\caption{Energy error versus iteration number for the new method,
within the conjugate gradient framework for simple cubic hydrogen when
using the Coulomb potential. The first few data points converge slowly
in order to accurately converge the fillings.}
\label{fig:Hconverge}
\end{figure}

\begin{figure}
\begin{center}
\mbox{\rotatebox{90}{{\hspace*{.3in}
      $log\left(\frac{(E-E_{min})}{\left| E_{min}\right| }\right)$}}
\scalebox{0.3}{\includegraphics{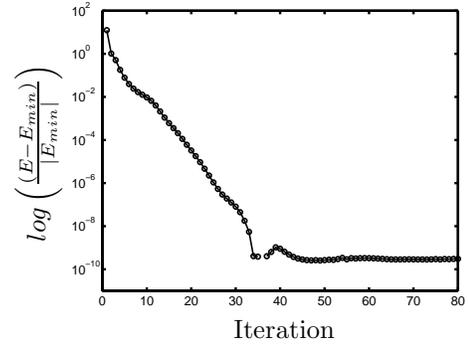}}} \\
{\hspace*{.4in} Iteration}
\end{center}
\caption{Energy error versus iteration number for the new method,
within the conjugate gradient framework for fcc carbon when using a
nonlocal pseudopotential.}
\label{fig:Cconverge}
\end{figure}

To underscore the effectiveness of conjugate-gradient methods for use
within our framework,
Figures~\ref{fig:Hconverge}~and~\ref{fig:Cconverge} show the iterative
convergence of the total energy within the local density approximation
(LDA)~\cite{perdew} to density-functional theory for cubic hydrogen
and an eight atom cell of fcc carbon, respectively.  The calculations
use the preconditioner of Teter, Payne and Allen~\cite{teter_cg},
cutoffs of $E^P_c = 8$~H and $E^Q_c = 16$~H for hydrogen and $E^P_c
=30$~H and $E^Q_c = 60$~H for carbon, and the overlap
operator $\mathcal{O}$ from Section~\ref{sec-details}.  The hydrogen
calculation uses the pure Coulomb potential and an $8\times 8\times 8$
Monkhorst-Pack~\cite{mp} sampling grid, and the carbon calculation
uses the Goedecker, Teter and Hunter pseudopotential~\cite{gth} with
simple $\Gamma$-point sampling.  Finally, the cubic lattice constants
were $2.760$~bohr and $6.746$~bohr for sc hydrogen and fcc carbon,
respectively.

It can be seen that the conjugate gradient technique is stable and
that fluctuations occur only in the final stages when the error is
$\Delta E/E_{min} \approx 10^{-8}$, well within acceptable
convergence. These fluctuations are associated with the fact that
$H^{QQ}_o$ is used instead of $H^{QQ}$.  In general, we have found
that the stability of the conjugate gradient does not depend on the
approximation for $C^Q$, but mainly depends on how and when we update
$C^Q$.
Table~\ref{table:energycutoff} compares the error in the energy per
atom with the new approach, with cutoffs stated above, to that with
the traditional approach, with cutoff $E^{trad}_c=E^P_c$. The fully
converged results are considered to be that of the traditional approach
with energy cutoff $E^{conv}_c=E^Q_c$.  The results for the new
procedure show a vast improvement.

As noted above, one of the advantages of our approach is that it
allows for an {\em a priori} estimate of the transferability errors.
We define this {\em transferability prediction} as
\begin{equation}
\Delta E^{predict}_{new} = \sum_{states}
\epsilon_{states}(\epsilon_{states}/E^P_c)^2, \label{eqn:epredict}
\end{equation}
where the sum is over occupied states.  For the specific cases of the
present two calculations, our prediction gives $ \Delta
E^{predict}_{new}\approx 0.5(0.5/8)^2 \approx 2mH$ for hydrogen and
$\Delta E^{predict}_{new} \approx 2\times 0.5(0.5/30)^2 \approx 0.3mH$
for carbon, where for simplicity we consider only the 2 s-states.
These predictions also appear in Table~\ref{table:energycutoff}. The
predicted values are very similar to $\Delta E_{new}$, demonstrating
that this prediction, which may be applied in {\em any} environment,
gives sensible estimates of the errors.

\begin{table}
\centering
\begin{tabular}{|l|c|c|c|}
\hline  & $\Delta E_{trad} (Ec=E^P_c)$ & $\Delta E_{new} $  & $\Delta E_{new}^{predicted}$ \\
\hline Hydrogen (sc)   & $9mH$  & $1mH$ & $2mH$\\
\hline Carbon (fcc)   & $16mH$ & $0.3mH$ & $ 0.3mH$\\
\hline
\end{tabular}
\caption{The error in calculating the energy per atom for simple cubic
hydrogen and fcc carbon.  The first column shows the error in energy
when using the traditional approach. The next column shows the error
in energy when using the new approach. The final column shows the
transferability prediction of the new approach,
Equation~(\ref{eqn:epredict}).}
\label{table:energycutoff}
\end{table}

Sections~\ref{sec-demos}~and~\ref{sec-boron} show that in practice the
present method is particularly beneficial for {\em all}-electron
calculations of first row elements, where the reduction in the cutoffs
is much more dramatic than the cases presented here.

\subsection{Application in the Analytic Continued Approach} \label{sec:analcont}

Our procedure can be easily incorporated into the analytic continued
conjugate gradient approach~\cite{arias_cg}. In such an approach,
constrained variables, $\{C\}$, are not minimized, but unconstrained
variables, $\{Y\}$, are minimized. This approach allows for much better
convergence, as all directions are allowed in the search space when
minimizing the energy functional~\cite{arias_cg}.

In the analytic continued approach, the sets $\{C\}$ and $\{Y\}$ are
related by
\begin{equation}
\bfC = \bfY u^{-\frac{1}{2}} V^{\dagger}, \label{eqn:CY}
\end{equation}
where bold faced quantities refer to the expansion coefficients for
all states gathered into matrices, each of whose columns represents a
particular state. Here, $V$ is a unitary transformation which allows
for subspace rotations~\cite{arias_cg,dftpp} and $u$ is the
expectation value of $\bfY$ with respect to the overlap matrix
$\mathcal{O}$,
\begin{equation}
u = \bfY^{\dagger}\mathcal{O}\bfY. \label{eqn:u}
\end{equation}
Note that Equation~(\ref{eqn:CY}) is constructed so that the $\bfC$
automatically satisfy the orthogonality constraint, with regard to
$\mathcal{O}$, as direct substitution verifies.  This approach allows
the total energy to be found by minimizing the energy $E[\bfY]$
directly using standard, unconstrained conjugate-gradient techniques.

In our case, the gradient becomes
\begin{eqnarray}
\frac{\partial E}{\partial \bfY^{P\dagger}} &= & (\identity -
\mathcal{O}\bfC^P \bfC^{P\dagger})(H^{PP} + H^{eff})
\bfC^PFVU^{\frac{1}{2}} \nonumber \\ & & + \ \mathcal{O}\bfC^P
VQ(V^{\dagger}[\tilde{H},F]V),
\label{eqn:gradYP}
\end{eqnarray}
where $H^{eff}$ is as defined in Equation~(\ref{eqn:Heff}), $F$ is a
diagonal ``filling'' matrix composed of the state occupancies,
$\mathcal{O}$ is the overlap matrix, defined similarly to
Equation~(\ref{eqn:Overlap}), and $Q()$ is the
$Q$-operator~\cite{arias_cg,dftpp}, defined by
$$
\left(W^{\dagger}Q(A)W\right)_{nm} \equiv
\frac{\left(W^{\dagger}AW\right)_{nm}}{\sqrt{\mu_n} + \sqrt{\mu_m}},
$$ 
where $W$ is the unitary, column order, eigenvector matrix for $u$
and $\mu$ are the corresponding eigenvalues.  In
Equation~(\ref{eqn:gradYP}), the $Q$-operator operates on the matrix
$V^{\dagger}[\tilde{H},F]V$, where $[\tilde{H},F]$ is the commutator
between the filling matrix $F$ and the subspace Hamiltonian matrix,
$$
\tilde{H} = \bfC^{P\dagger}(H^{PP} + H^{eff})\bfC^P.
$$
Now, $u$ is redefined as
\begin{equation}
u = \bfY^{P\dagger}\mathcal{O}\bfY^P \label{eqn:up}
\end{equation}
and $\bfC^P$ is related to $\bfY^P$ by
$$
\bfC^P = \bfY^P u^{-\frac{1}{2}} V^{\dagger}.
$$
Finally, the solution for
$\bfC^Q$ becomes
\begin{equation}
\bfC^Q = -\frac{1}{H^{QQ}_o}H^{QP}\bfC^P.
\label{eqn:CQCG}
\end{equation}

We note that $H^{eff}$ in Equation~(\ref{eqn:gradYP}) acts as a
pseudopotential generated directly from the crystal environment and
not from a referenced atomic calculation which one must hope to be
transferable. The overlap matrix $\mathcal{O}$ acts to preserve the
correct orthonormalization for the $P$-space wave function.  This
overlap matrix is similar to those found in the Ultra-Soft
Pseudopotentials~\cite{ussp} and Projected Augmented
Wave~\cite{paw} methods. In those methods, the softening procedure is
generated in a real space formalism, from a specific reference state,
and depends on the choice of a cutoff in real space cutoff. In this
work, the softening is done in a momentum space formalism without
artificial core radii, making the procedure ideal for high-pressure
studies where the loss of distinction between core and valence regions
is precisely the physics of interest.  As a final advantage of the
present approach, Equation~(\ref{eqn:epredict}) gives a direct measure
of transferability errors.

\section{Computational details}~\label{sec-details}

We will now describe briefly efficient calculation of specific terms
in our procedure.  The text focuses on all-electron calculations as
these are of interest in the present work.  The appendix discusses
details for the use of this approach with hard but transferable
Kleinman-Bylander pseudopotentials.

\subsection{Form for Overlap Matrix}

The overlap matrix has the form
\begin{equation}
\mathcal{O} = \identity + \left( H^{PQ}_{h,xc} + H^{PQ}_{ion}\right)
\frac{1}{H^{QQ}_o} \frac{1}{H^{QQ}_o} \left( H^{QP}_{h,xc} +
H^{QP}_{ion}\right),
\label{eqn:over_decomp}
\end{equation}
where $\frac{1}{H^{QQ}_o} $ is the inverse of the kinetic energy,
$H_{ion}$ is the contribution from the local ionic potential and
$H_{h,xc}$ is the contribution from the Hartree and
exchange-correlation potentials, which depends on the density.  There
are a number of viable options for implementing $\mathcal{O}$.  We
have explored both the option of fixing $H_{h,xc}$, and thus
$\mathcal{O}$ to some value and the option of updating it
self-consistently.

The benefit of holding the overlap matrix fixed is that 
\begin{equation}
\frac{\partial \mathcal{O}}{\partial \bfC^{P\dagger}} = 0, \label{eqn:dO}
\end{equation}
so that Equation~(\ref{eqn:gradPold}) is the exact variational
derivative of the energy functional.  The appendix further shows that
such a fixed operator makes efficient implementation of
Kleinman-Bylander pseudopotentials possible.  A clear choice for a
fixed overlap matrix is to hold $H_{h,xc}$ in
Equation~(\ref{eqn:over_decomp}) to that of the ``free-atom'' crystal,
a crystal whose charge density is just the superposition of the charge
densities from isolated atoms.  In such a case, the overlap matrix
becomes
\begin{eqnarray}
\mathcal{O} &=& \identity + H^{cr-at,PQ} \frac{1}{H^{QQ}_o} \frac{1}{H^{QQ}_o}
H^{cr-at,QP},
\label{eqn:overat}
\end{eqnarray}
where $ H^{cr-at,PQ} = H^{cr-at,PQ}_{h,xc} + H^{PQ}_{ion}$ and
$H^{cr-at,PQ}_{h,xc}$ is Hartree and exchange-correlation potential
for the ``free-atom'' crystal.  One possible cause for concern when
using a fixed overlap matrix is that the orthonormalization is no
longer exact (so that, for instance, the electronic charge will not be
precisely conserved) because $H^{QP}$ in Equation~(\ref{eqn:CQCG})
will differ from its free-atom crystal value.  However, we have found
this to never affect the convergence of the calculations and to yield
quite accurate results, as
Figures~\ref{fig:Hconverge}~and~\ref{fig:Cconverge} and
Table~\ref{table:energycutoff} show.

Alternately, exact orthonormalization can be preserved when using a
fixed overlap operator if one chooses to redefine $\bfC^Q$ as
\begin{equation}
\bfC^Q =
-\frac{1}{H^{QQ}_o}H^{cr-at,QP}\bfC^P.
\label{eqn:CQat}
\end{equation}
While this preserves orthonormality among the states, it does give a
slightly different approximation to $\bfC^Q$, which causes a slight
change in the $H^{eff}$ which should be used in (\ref{eqn:ham_eff}).
We find, however, that ignoring this discrepancy makes no practical
difference in the final results.  With this alternate approach, the
convergence is quite similar to those of
Figures~\ref{fig:Hconverge}~and~\ref{fig:Cconverge}, and the energy
per atom is within $1\mu$~H, as compared to using
Equation~(\ref{eqn:CQCG}) for $\bfC^Q$. Another benefit of using the
form Equation~(\ref{eqn:CQat}) comes when using non-local
pseudopotentials, as this form is computationally more efficient.
(See the appendix for details.)

Finally, the $H^{eff}$ which should be used with the construction in
Equation~(\ref{eqn:CQat}) is
\begin{equation}
H^{eff} = -H^{PQ}\frac{1}{H^{QQ}_o}H^{cr-at,QP}. \label{eqn:Heffat}
\end{equation}
Maintaining this consistency comes at the cost of introducing a
slightly non-Hermitian Hamiltonian.  Despite this, we again have not
experienced any practical problems working with this form.

The second option for implementing the overlap matrix is to use the
Hartree and exchange correlation potentials self-consistently. When
doing so, the gradient of the total energy involves terms including
the derivative of the overlap matrix.  Unfortunately, we have yet to
find a method for calculating these contributions to the gradient
which does not require multiplication by matrices of size $N^P\times
N^P$, a completely impractical feat for all but the most trivial of
systems.

\begin{figure}
\begin{center}
\rotatebox{90}{ {{\hspace*{.3in}  $log\left(\frac{(E-E_{min})}{\left|
	E_{min} \right|}\right)$}}}
\scalebox{0.3}{\includegraphics{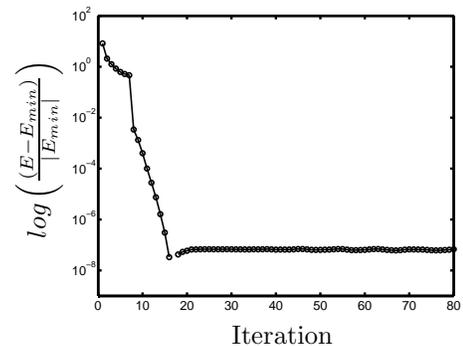}} \\ {
\hspace*{.4in} Iteration}
\end{center}
\caption{Energy convergence versus iteration for simple cubic hydrogen
when the overlap matrix is updated self-consistently, but its
variational derivative is not taken into account.}
\label{fig:Hconverge_Osc}
\end{figure}

\begin{figure}
\begin{center}
\rotatebox{90}{{ {\hspace*{.3in}  $log\left(\frac{(E-E_{min})}{\left|
	E_{min}\right| }\right)$}}}
\scalebox{0.3}{\includegraphics{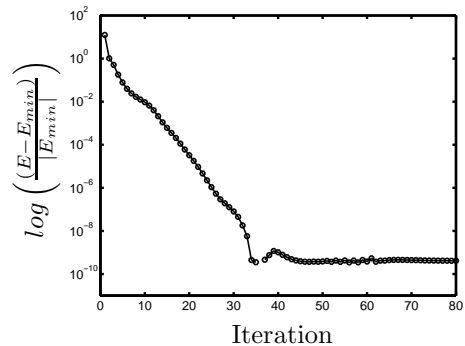}} \\
{\hspace*{.4in} Iteration}
\end{center}
\caption{Energy convergence versus iteration for fcc carbon when the
overlap matrix is updated self-consistently, but its variational
derivative is not taken into account. }
\label{fig:Cconverge_Osc}
\end{figure}

We have found somewhat to our surprise, however, that updating the
overlap matrix self-consistently while completely ignoring these
contributions to the gradient, that the gradient technique remains
stable and maintains its good convergence, as
Figures~\ref{fig:Hconverge_Osc}~and~\ref{fig:Cconverge_Osc} show.  In
these calculations, $H^{cr-at}_{h,xc}$ was used only at the first
iteration to initialize the calculation.  After that, the overlap was
updated self-consistently and the gradient was always calculated
ignoring the aforementioned contributions from the changes in
$\mathcal{O}$.  The convergence using this technique is quite similar
to that which holds $\mathcal{O}$ fixed. The only noticeable
difference occurs in the asymptotic region, where now slightly larger
fluctuations exist as fluctuations in the overlap matrix are not
properly taken into account.  Finally, we find the final energy per
atom again to be within $1\mu$H of all of the approaches above.

\subsection{Numerical Implementation}

Specifying the mathematical form of the overlap matrix in one of the
forms from the previous section and thus also the forms for $\bfC^Q$
and $H^{eff}$ leaves the task of evaluating these and related quantities
numerically.  This section first discusses efficient methods for this
evaluation and then describes how to build conjugate gradient
minimization from the ability to perform these evaluations.

The term $\mathcal{O}\bfY^P$ must be calculated for defining $u$,
Equation~(\ref{eqn:up}), and in the gradient,
Equation~(\ref{eqn:gradYP}). The most efficient method for calculating
this term is to perform four Fourier transforms per electronic
wave function, always multiply terms in the space in which they are
diagonal.  We find it useful to keep this term in storage so as to
only calculate it once in a conjugate gradient loop. Such storage is
minimal, as will be seen.

To reduce the time spent in Fourier transforms, we make the
approximation of using a slightly smaller Fourier transform grid than
might normally be used.  The gradient, the high momentum coefficients
$\bfC^Q$, and the overlap matrix only involve momentum transfers of
$P+Q$.  Higher momentum transfers do not occur because $H^{QQ}$ is
approximated by $H^{QQ}_o$, which is diagonal.  Accordingly, rather
than Fourier grids of size $2Q$, we employ grids of size $P+Q$.  We
find that using the full Fourier transform grid decreases the energy
minimally: for hydrogen and carbon doing so decreases the energy per
atom by only $0.5\mu$H and $0.8\mu$H, respectively.

The reduction of the Fourier transform grid to size $P+Q$ from a size
of $2Q$, which would be needed in a non-perturbative approach, saves
significant storage in the ratio
\begin{equation}
\frac{FFT_{new}}{FFT_{full}} = \frac{1}{8}\left(1 + 3
\left(\frac{E_P}{E_Q}\right)^{\frac{1}{2}} + 3
\left(\frac{E_P}{E_Q}\right) +
\left(\frac{E_P}{E_Q}\right)^{\frac{3}{2}} \right).   \label{eqn:fftsave}
\end{equation}
In our boron calculation, this is a factor of nearly three,
which can make the difference between being able to perform the
calculation on a workstation or requiring parallel computing capability.

We find the following procedure to be very stable for performing
conjugate gradient minimization.  First, given $\bfC^P$ generate
$\bfC^Q$ according to
Equation~(\ref{eqn:CQCG})~or~(\ref{eqn:CQat}). Next, calculate the
total energy and the gradient for $\bfY^P$ at this point from
Equation~(\ref{eqn:gradYP}). Holding $\bfC^Q$ fixed, minimize $\bfY^P$
along the search direction, and generate a new $\bfY^P$, and iterate
to convergence.

This procedure for solving for the $P$-space and generating the
$Q$-space, when needed, is very beneficial for plane-wave calculations
when using either the Coulomb potential or a very hard, yet highly
transferable, pseudopotentials because, generally for such
calculations the computational bottleneck is memory and this approach
requires permanent storage only in the $P$-space.  (The $Q$-space wave
functions for a particular band can be generated only when needed.)
Moreover, in terms of computational time, a savings of $1 + N_Q/N_P$
for all direct matrix multiplications occurs, and the $P$-space
approach is therefore also more efficient in terms of computational
time for systems with many bands.

\subsection{Application to the Coulomb Potential}\label{sec-demos}

To demonstrate the benefits of our approach, this section presents
all-electron calculations for fcc boron at two different volumes
($V_1=3.83$~\AA$^3$ and $V_2=3.5$~\AA$^3$), exploring convergence with
energy cutoff, usage of memory and computational time, and our ability
to predict transferability errors.  We carry out all calculations
within the local density approximation with ten special k-points in
the reduced Brillouin zone and a Fermi surface integration temperature
of $kT \approx 0.0037$~H.

\begin{figure}
\begin{center}
\mbox{\rotatebox{90}{\hspace*{.6in} Energy per Atom [eV] }
\scalebox{0.45}{\includegraphics{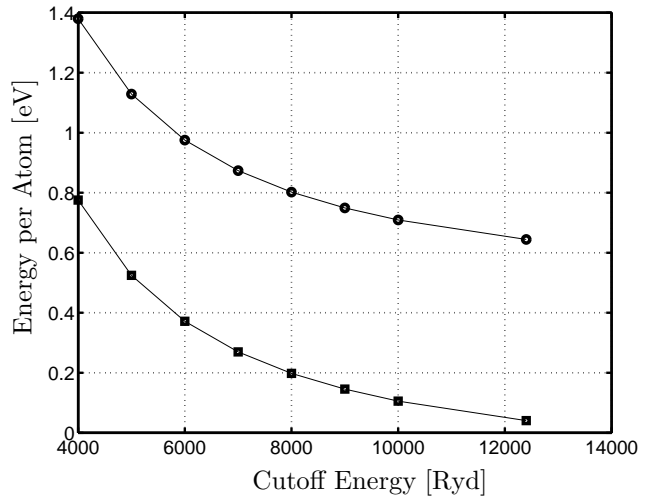}}} \\
{\hspace*{.4in} Cutoff Energy [Ryd]}
\end{center}
\caption{Energy per atom as a function of cutoff energy for fcc boron,
when using the Coulomb potential in the traditional plane-wave
approach. The squares correspond $V_1=3.83$~\AA$^3$ and the circles
correspond to $V_2 = 3.5$~\AA$^3$. }
\label{fig:converge}
\end{figure}

\begin{figure}
\begin{center}
\mbox{\rotatebox{90}{\hspace*{.6in} Energy Difference [eV] }
\scalebox{0.45}{\includegraphics{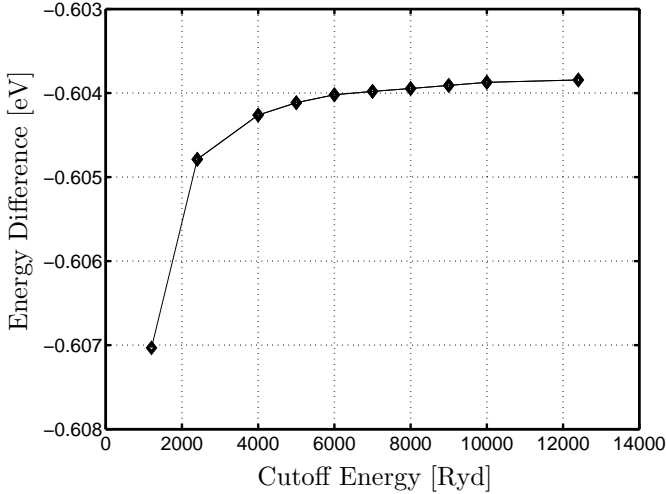}}} \\
{\hspace*{.4in} Cutoff Energy [Ryd]}
\end{center}
\caption{Energy difference as a function of cutoff for fcc boron, when
using the traditional plane-wave approach. The volumes per atom are 
$V_1=3.83$~\AA$^3 \mbox{ and } V_2=3.5$~\AA$^3$.  }
\label{fig:ediff}
\end{figure}

Figure~\ref{fig:converge} shows the convergence of the energy as a
function of cutoff for the direct plane wave approach.  The data in
the figure demonstrate that the traditional plane-wave approach
converges quite slowly, requiring an energy cutoff of over 12000 Ryd
to bring the total energy to within 0.1~eV per atom.  However, as is
well known, the higher-energy plane wave states provide convergence
mostly to the inert region of the core near the nucleus and thus
physically meaningful energy differences converge much more rapidly
than the absolute total energy.  As a more meaningful reference,
Figure~\ref{fig:ediff} shows the convergence of the energy difference
between volume states $V_1$ and $V_2$ using the direct plane wave approach.

\begin{figure}
\begin{center}
\mbox{\rotatebox{90}{\hspace*{.6in} Energy per Atom [eV] }
\scalebox{0.45}{\includegraphics{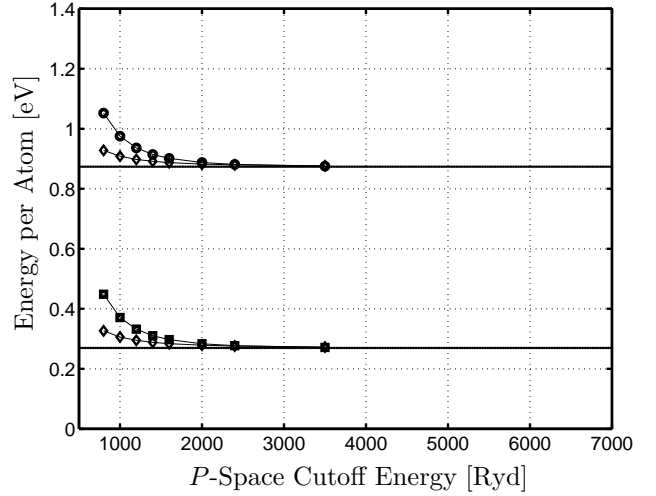}}} \\
{\hspace*{.3in} $P$-Space Cutoff Energy [Ryd]}
\end{center}
\caption{Energy convergence versus $P$-space cutoff, using the new
method, when $E^Q_c = 7000$ Ryd. The squares correspond to
$V_1=3.83$~\AA$^3$, the circles correspond to $V_2 = 3.5$~\AA$^3$ and
the diamonds correspond to the transferability prediction,
Equation~(\ref{eqn:epredict}). The horizontal lines correspond to the
energy calculated using the traditional approach at a cutoff energy of
7000 Ryd.}
\label{fig:convergeps}
\end{figure}

Figure~\ref{fig:convergeps} shows the convergence with $P$-space
cutoff of the total energy when using the perturbative approach with a
$Q$-space cutoff of 7000 Ryd and employing $H^{cry-at}_{h,xc}$ for
construction of $\mathcal{O}$, $\bfC^Q$ and $H^{eff}$
(Equation~(\ref{eqn:Heffat})).  The figure shows that there is almost no
difference in the total energy predictions (circles and squares for
the two volume states) when reducing the cutoff in the $P$-space to
3500 Ryd and that reducing the $P$-space cutoff down to 1000 Ryd gives
just as accurate total energies as a 6000 Ryd cutoff in the
traditional approach.  The figure also shows that the transferability
prediction based on the two core electrons (diamonds) gives the
correct order of magnitude and is never off by more than a factor of
three in this case.

\begin{figure}
\begin{center}
\mbox{\rotatebox{90}{\hspace*{.6in} Energy Difference [eV] }
\scalebox{0.45}{\includegraphics{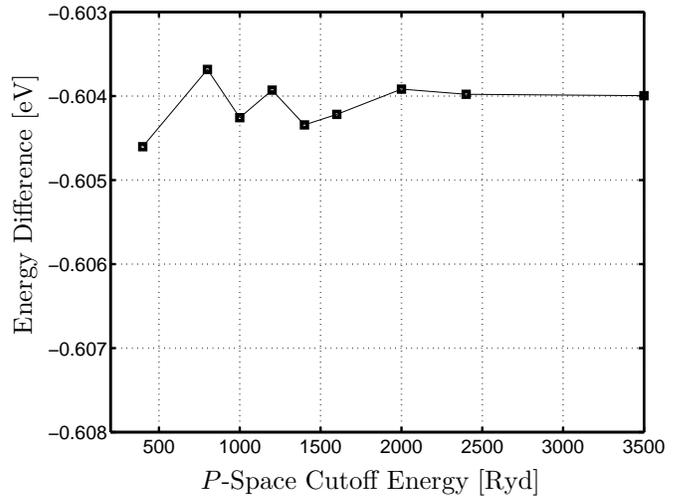}}} \\
{\hspace*{.3in} $P$-Space Cutoff Energy [Ryd]}
\end{center}
\caption{Energy difference in the new approach as a function of the
$P$-space cutoff.  }
\label{fig:ediffps}
\end{figure}

Figure~\ref{fig:ediffps} shows the performance of the new approach in
computing energy differences.  Energy differences in the perturbative
approach tend to fluctuate slightly and lack the monotonic behavior
which the traditional approach exhibits due to loss of ability to
describe physical changes in the high momentum states.  Perturbation
theory recovers these changes nearly perfectly and so the resulting
error in energy difference is smaller and loses systematic behavior.
From these results we conclude that a cutoff in the $P$-space of 1200
Ryd and of 7000 Ryd in the $Q$-space is more than sufficient to
predict accurate energy differences and to capture any unforeseen
physical processes which come into play.

To make a direct comparison of computational savings, we note that the
total energy and energy difference, when using the new method with
$E^P_c$ = 1200 Ryd and $E^Q_c$ = 7000 Ryd, is very close to the
traditional approach with energy cutoff $E^{trad}_c$ = 6500 Ryd. At
these cutoffs, the memory savings from the perturbative approach is a
factor of 12.5 for the wave functions and 2.7 for the FFT's.  The time
for matrix multiplications decreases by a factor of 5.4 and for
Fourier transforms by a factor of 2.7 which largely compensates the
need for a few more Fourier transforms in generating $\bfC^Q$ and
$\mathcal{O}\bfY^P$.

\section{Application to High Pressure Boron}\label{sec-boron}

As a physical application, we study the high-pressure phases of boron,
whose rich physics remains mysterious.  Boron is known to exhibit a
semiconducting to metallic phase transition at $\approx
170$~GPa\cite{eremets}, and under high pressure and low temperatures,
boron superconducts~\cite{eremets}.  Previous theoretical studies of
boron under pressure include that of Mailhiot {\em
et. al.}\cite{mailhiot}, who used both the LMTO method and the
pseudopotential method.  They have found their pseudopotential results
to be more reliable and predict a sequence of structural phase
transitions with increasing pressure from the icosahedral structure
(\a12) to a body centered tetragonal structure (bct) to the
face-centered cubic structure (fcc).  They find the \a12 $\rightarrow$
bct transition to occur at 210 GPa, and established this as an upper
bound for the semiconducting to metallic phase transition, within the
local density approximation.

To calculate the energy for different phases in boron we use our
perturbative potential approach with the analytically continued
conjugate gradient approach as described in Section~\ref{sec-cg} with
$\mathcal{O}$, $\bfC^Q$ and $H^{eff}$ calculated as in
Section~\ref{sec-demos}.  All calculations employ the local density
approximation with the parameterization of Perdew-Zunger~\cite{perdew}
to the Ceperley-Alder~\cite{ceperley} exchange-correlation energy.

We also consider other structures than explored
previously~\cite{mailhiot,zhao}, not only fcc, bct, and \a12
structures but also orthorhombic structures of boron with the
$\alpha$-gallium and $\beta$-gallium basis, which we denote as \aortho
and \mbox{$\beta_{ga}$-B,} respectively.  We consider these
orthorhombic structures for a number of reasons.  Zhu and
Henley~\cite{zhu} have shown that for 6-coordinated boron under high
pressure the triangular lattice is a good candidate for a low energy,
high pressure phase, within the density functional theory
pseudopotential framework~\cite{payne} and furthermore such sheets
tend to buckle~\cite{zhu,boustani1,boustani2}. Because the gallium
structure can be viewed as stacked sheets of a highly buckled
triangular lattice, such a structure is a good candidate for a high
pressure phase of boron. Secondly, gallium is another Group III
element which has as its ground-state structure $\alpha$-gallium at
low pressures and $\beta$-gallium at high pressures\cite{donohue}, and
being under boron in the periodic table supports the notion that such
structures are good candidates for a high pressure phase of boron.
Below, we show that these structures play quite important roles.

For the bct structure, we use the same $c/a = 0.65$ ratio as was used
in Reference~\onlinecite{mailhiot}, where this ratio was shown to give the
ground state at a volume $\approx 3.91$ \AA$^3$ within the
pseudopotential calculation. We have tested this ratio for a number of
high pressure states and have found the ideal ratio to remain constant
over a wide range of pressures.  Hence, we hold $c/a$ at this value
for all calculations in our study, and thus our bct energies results
represent a quite close upper-bound.

For the \a12 structure, we use the structure listed in Table 5 of
Reference~\onlinecite{donohue}, the same used by Mailhiot {\em
et. al.}~\cite{mailhiot}.  This structure is rhombohedral with angle
$\alpha = 58.06^\circ$.  The basis locations for the atomic positions
are
$$
\pm (x_1x_1z_1; \ x_1z_1x_1; \ z_1x_1x_1; \ x_2x_2z_2; \ x_2z_2x_2; \
z_2x_2x_2),
$$
where $x_1 = 0.0104, \ z_1 = -0.3427, \ x_2 = 0.2206 \mbox{ and } z_2
= -0.3677$.  The space group for the \a12 structure is R$\bar 3$m.
Note that we may regard this \a12 structure as representative of the
low-pressure states of boron which are all governed by the icosahedron.

For the orthorhombic structures, we use the Gallium structures as
found in Wyckoff's book~\cite{wyckoff}. For the \aortho structure, we
hold the lattice vector ratios to $a/b = 0.99867$ and $a/c =
0.590035$ and place the atomic basis at
coordinates
$$
\pm (0,u,v; \ 1/2,u+1/2,\bar v; \ 1/2,u,v+1/2; \ 0,u+1/2,1/2-v),
$$ where $u = 0.0785$ and $v = 0.1525$. The space group is Cmca. For
the \bortho structure, we hold the the lattice vector ratios at $a/b =
0.3567$ and $a/c = 0.9148$ and place the atomic basis at
$$
\pm (0,u,1/4; \ 1/2,u+1/2,1/4),
$$ where $u = 0.133$. The space group is Cmcm.  

Note that we do not optimize the lattice vector ratios or the internal
coordinates for the orthorhombic structures for boron but here hold
them fixed to that of gallium at standard conditions.  Therefore, the
energies which we report for these structures should be regarded
simply as upper bounds.  Similarly, we do not optimize the lattice
vectors and/or internal coordinates for the bct and \a12 structures,
however this latter optimization proves to be minor.  Zhao and
Lu~\cite{zhao} have recently performed pseudopotential calculations,
for the fcc, bct and \a12 structures, and found that fully relaxing
these structures only decrease the transition pressure (bct
$\rightarrow$ \a12) by 5\%, when using the local density
approximation, as compared to Reference~\onlinecite{mailhiot}. Therefore,
because these latter effects are minor near the
semiconducting/metallic transition pressure and in order to allow our
method a greater comparison with pseudopotential results we choose to
hold the parameters for the bct and \a12 structures fixed to those
used in Reference~\onlinecite{mailhiot}.

\subsection{Convergence}

We now establish plane wave cutoffs for our specific study which will
both make for practical calculations (particularly for the \a12
structure) and give accurate structural energy differences.  For this
we shall compare the energy per atom as a function of volume for the
fcc, bct and \a12.  As these are convergence tests,
we employ relatively modest Brillouin sampling: 4$\times$4$\times$4,
4$\times$4$\times$6, and 1$\times$1$\times$1 (the $\Gamma$ point)
meshes for the fcc, bct and \a12, respectively.  To
facilitate integration over the Fermi surface, we employ an electronic
temperature of $kT \approx 0.0037$~H.

\begin{figure}
\begin{center}
\mbox{\rotatebox{90}{\hspace*{.6in} Energy  per Atom [eV] }
\scalebox{0.45}{\includegraphics{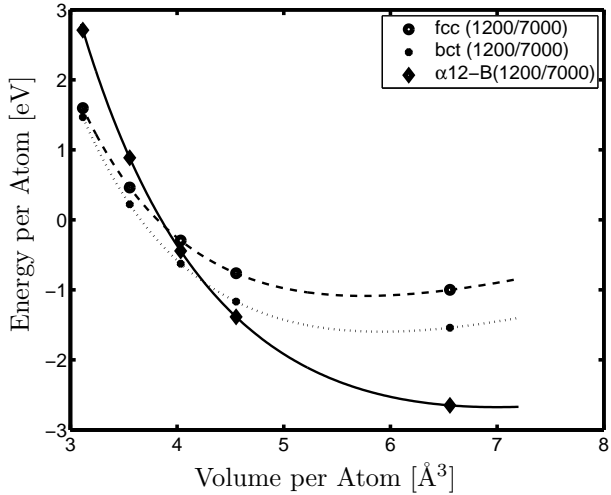}}} \\
\hspace*{.1in} Volume per Atom [\AA$^3$]
\end{center}
\caption{Energy per atom versus volume per atom for the fcc (circles),
bct (stars) and \a12 (diamonds) structures. Energy cutoff for the
$P$-space is 1200 Ryd and for the $Q$-space 7000 Ryd. The curves are
fitted to Equation~(\ref{eqn:expansion}). }
\label{fig:EvsV_1200_7000}
\end{figure}

Figure~\ref{fig:EvsV_1200_7000} summarizes the results for the three
structures when using the plane wave cutoffs from
Section~\ref{sec-demos}, $E^Q_c$ = 7000 Ryd and $E^P_c$ = 1200 Ryd.
The curves from the figure are simple fits to
\begin{equation}
E = a_0 + a_1 V^{-1/3} + a_2 V^{-2/3} + a_3 V^{-1}.
\label{eqn:expansion}
\end{equation}

\begin{figure}
\begin{center}
\mbox{\rotatebox{90}{\hspace*{.6in} Energy per Atom [eV] }
\scalebox{0.45}{\includegraphics{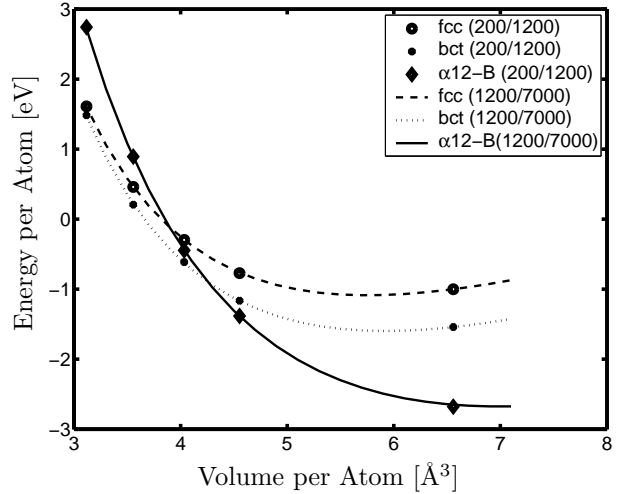}}} \\
{\hspace*{.1in} Volume per Atom [\AA$^3$]}
\end{center}
\caption{Energy per atom versus volume per atom. The results from new
method using the cutoffs $E^P_c$=200 Ryd and $E^Q_c$=1200 Ryd are
displayed as circles (fcc), stars (bct) and diamonds
(\a12). The results from new method using the cutoffs
$E^P_c$=1200 Ryd and $E^Q_c$=7000 Ryd are displayed as lines by fitting
the data to Equation~(\ref{eqn:expansion}): fcc (dashed), bct
(dotted), \a12 (solid).}
\label{fig:EvsV_200_1200}
\end{figure}

Having established previously that the above cutoffs give extremely
good energy differences, we then searched for possible reductions in
the cutoffs which continue to reproduce these results accurately.  We
find reducing the cutoffs to 200 Ryd for the $P$-space and 1200 Ryd
for the $Q$-space to both lead to efficient calculations and to give
highly accurate results.
Figure~\ref{fig:EvsV_200_1200} compares the results at this reduced
cutoff (data points) with the previous results at the higher cutoffs
(curves).

\begin{figure}
\begin{center}
\mbox{\rotatebox{90}{\hspace*{.6in} Energy per Atom [eV] }
\scalebox{0.45}{\includegraphics{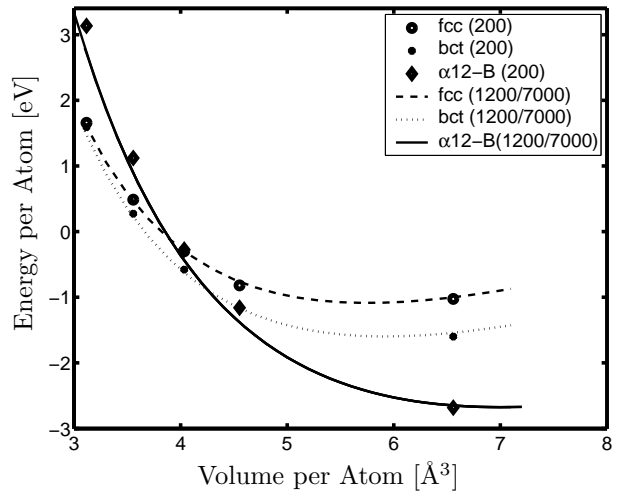}}} \\
{\hspace*{.1in} Volume per Atom [\AA$^3$]}
\end{center}
\caption{Energy per atom versus volume per atom. The results from
using the traditional approach with cutoff $E$=200 Ryd are displayed
as circles (fcc), stars (bct) and diamonds (\a12).  The results
from new method using the cutoffs $E^P_c$=1200 Ryd and $E^Q_c$=7000
Ryd are displayed as lines by fitting the data to
Equation~(\ref{eqn:expansion}): fcc (dashed), bct (dotted), \a12
(solid).}
\label{fig:EvsV_200}
\end{figure}

As a comparison, we also calculate the energy when
using a traditional, direct plane-wave approach with a cutoff of 200
Ryd. Figure~\ref{fig:EvsV_200} shows a similar plot, where the points
are the plane wave data and the curves represent the fit to the
fully converged results.  Note that, although
fcc structure is fairly well converged at this low cutoff, there
are rather large errors for the \a12 structure.  Thus, such a low
cutoff is unreliable for the study of such systems.
The fact the low cutoff represents one structure well and not another
demonstrates the unpredictability of transferability errors,
underscores the need to predict those errors reliably, and places
pseudopotential studies of such systems, which assume that errors in
processes at such high energy scales cancel, in doubt.

Finally, we note that the memory savings of using the new method, with
cutoffs of 200 Ryd and 1200 Ryd, versus the traditional approach,
which would require a cutoff of 1200 Ryd to produce results of similar
reliability, is quite substantial, a factor of $\approx 15$ for wave
functions and $\approx 3$ for the FFT box. The time for the matrix
multiplication correspondingly reduces by a factor of 6 and for the
Fourier transforms by a factor of 3, similar to
Section~\ref{sec-demos}.  Because of these savings, particularly in
memory, all of our calculations below were possible on a single
desktop computer and there was no need for parallel supercomputing.

\subsection{Results}

Having found appropriate plane wave cutoffs, we next converged both
the Brillouin sampling and the fictitious Fermi temperature.  We find
that a fictitious Fermi temperature of $kT=0.001 H$ converges the
total energy to a within a few tenths of a millihartree per atom ($0.2
mH$ for the fcc structure).  For the fcc and bct structures, Brillouin
sampling on 6$\times$6$\times$6 and 4$\times$4$\times$6 meshes,
respectively, suffice to converge the total energy per atom to within
the same tolerance.  We then generate meshes for the remaining
structures by maintaining the same reciprocal space sampling density
as closely as possible, resulting in meshes of size
2$\times$2$\times$2, 4$\times$4$\times$4 and 6$\times$4$\times$6 for
the \a12, \aortho and \bortho structures, respectively.

\begin{figure}
\begin{center}
\mbox{\rotatebox{90}{\hspace*{.6in} Energy per Atom [eV] }
\scalebox{0.45}{\includegraphics{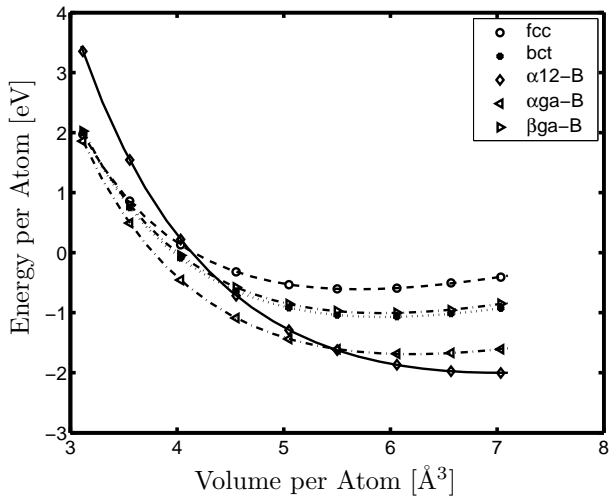}}} \\
{\hspace*{.1in} Volume per Atom [\AA$^3$]}
\end{center}
\caption{Energy per atom versus volume per atom, for the fcc
(circles), bct (stars), \a12 (diamond), \aortho (left facing
triangles) and \bortho (right facing triangles) structures.  The
results are from the new method using the cutoffs $E^P_c$=200 Ryd and
$E^Q_c$=1200 Ryd.}
\label{fig:EvsV}
\end{figure}

Figure~\ref{fig:EvsV} shows our final results, which are converged to
within a few tenth of a millihartree per atom.  Most notably, we find
a new phase, the \aortho, to be the lowest in energy phase under high
pressure. \aortho boron becomes energetically favorable over \a12
boron at a pressure of 71 GPa, and remains the energetically favored
phase up to the highest pressures we have studied, which are over 500
GPa, where the pressure is determined from the enthalpy.
Although there are other stable structures governed by the icosahedron
in boron, due to the dramatic rise in energy of the \a12 structure, we
do not expect these alternate structures to have significantly lower
energies. Moreover, Zhao and Lu~\cite{zhao} have preformed preliminary
pseudopotential calculations with the $\beta$-B$_{105}$ unit cell,
part of the icosahedral family, and find similar trends between this
structure and the \a12 structure.  Finally, optimizing the lattice
vectors ratios for the orthorhombic phases will only make them more
favorable.  We, therefore, expect our prediction of a phase transition
from the icosahedral family to the \aortho structure to be robust.

As a point of comparison to the pseudopotential
calculation\cite{mailhiot}, which had not considered the orthorhombic
structures, ignoring those structures, we also find the sequence of
transitions \a12 $\rightarrow$ bct $\rightarrow$ fcc.  Our value of
194~GPa for the first of these transitions agrees reasonably well with
the pseudopotential result of 210~GPa.  However, our prediction of
$\approx 475$~GPa for the second transition differs substantial from
the pseudopotential result of 360~GPa.  These discrepancy most likely
results from the inadequacies of the pseudopotential to effectively
account for the response within the core.  The results in
Figure~\ref{fig:EvsV_200} demonstrates the importance of momentum
transfers over 200~Ryd and hence variations $<$ 0.44 a$_0$, where
a$_0$ is the Bohr radius, well within the core region.  This
underscores the importance of control over transferability and
properly accounting for the core region when studying high pressure
systems.

\begin{figure}
\begin{center}
\mbox{\rotatebox{90}{\hspace*{.7in} Density of States }
\scalebox{0.45}{\includegraphics{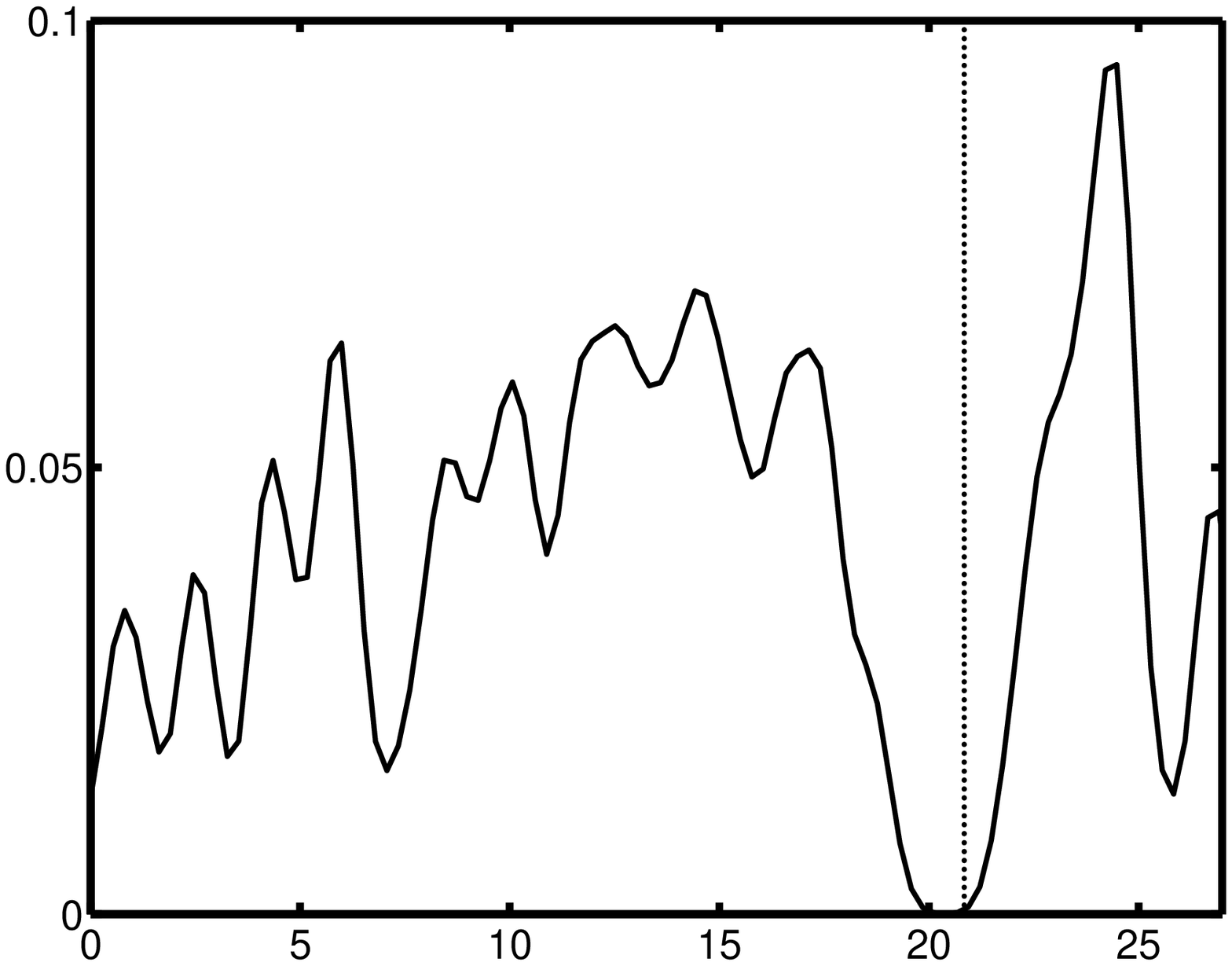}}} \\
{\hspace*{.1in} level [eV]}
\end{center}
\caption{Density of states for the valence electrons for the
\aortho  structure. The straight line corresponds
to the chemical potential at $kT=0.001H.$}
\label{fig:ds_alpha}
\end{figure}

\begin{figure}
\begin{center}
\mbox{\rotatebox{90}{\hspace*{.7in} Density of States }
\scalebox{0.45}{\includegraphics{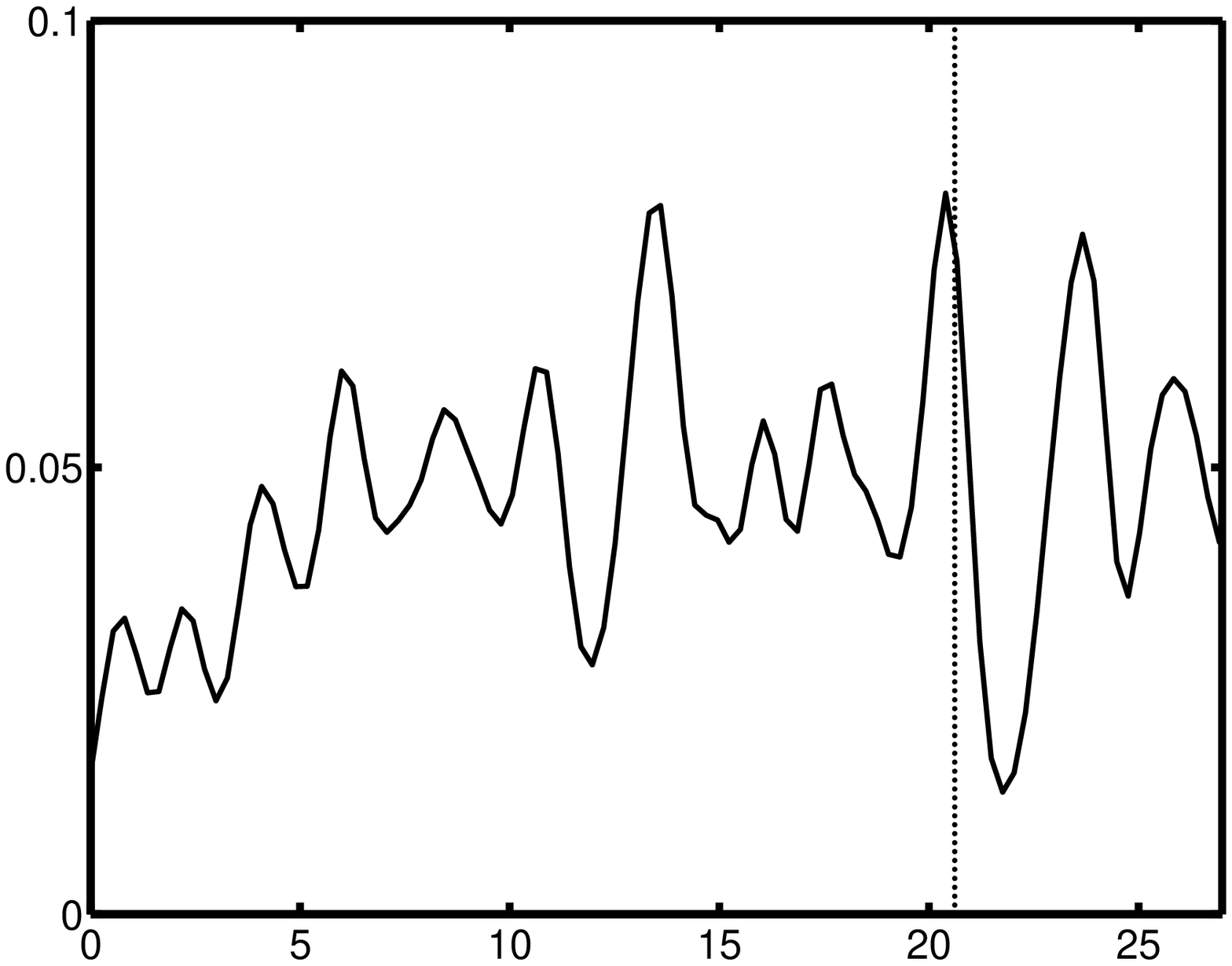}}} \\
{\hspace*{.1in} level [eV]}
\end{center}
\caption{Density of states for the valence electrons for the
\bortho structure. The straight line corresponds to
the chemical potential at $kT=0.001H$.}
\label{fig:ds_beta}
\end{figure}

Given the availability of high-pressure resistivity data for
boron\cite{eremets}, it is interesting to consider the electronic
structure of the hitherto unexplored \aortho phases for boron.
Figures~\ref{fig:ds_alpha}~and~\ref{fig:ds_beta} show low-resolution
plots of the density of states for \aortho and \bortho boron at
250~GPa and 217~GPa, respectively, as generated using the Brillouin
zone sampling from the total energy calculations.  The plots have low
resolution because, as is well known, meshes which give reliable total
energies often are too modest to give detailed density of states.
(Providing smooth curves required a Gaussian broadening of width
1~eV.)  The figures clearly show that, whereas the \bortho phase is
semimetallic, the density of states of the \aortho
phase clearly exhibits a gap at the Fermi level, even at high
pressures.

Experimentally, the room-temperature resistance of boron decreases
discontinuously as a function of pressure at 30~GPa, 110~GPa and
170~GPa\cite{eremets}.  In comparing our results to these data, it is
important to recall that the \a12 structure is only meant to be
representative of the icosahedral family of low-pressure structures
and that, with optimization of the aspect ratios and internal
coordinates, the curves for the \aortho and \bortho structures will
displace to even lower energies.  In particular, a downward shift of
only 0.15~eV (~10\%) in the \aortho curve would lower our prediction
for the \a12$\rightarrow$\aortho transition from 71~GPa to 30~GPa.
Thus, while further investigation is clearly needed, these data
suggest that a structural transition from the icosahedral family to
the orthorhombic phase is a viable candidate for the resistance
discontinuity observed in boron at 30 GPa. It is also interesting to
note that the \aortho structure (Cmca space group) was also observed
to be stable low energy, high pressure phases for both lithium and
sodium~\cite{neaton1,neaton2}, within density functional theory.

Because boron is observed experimentally to be metallic above 160~GPa,
our results imply that another, yet unexplored, structure must become
energetically favored at this pressure and be responsible for the
highest of the observed discontinuities in the room-temperature
resistance.  Whatever this metallic structure is, it cannot be simple
as the fcc, bct, bcc~\cite{mailhiot} and ideal hcp~\cite{mailhiot}
structures because none are competitive at high pressures.  It is
suggestive, however, that our energetic upper-bound for the metallic
\bortho structure is quite competitive with the fcc and bct structures
at the very highest pressures.  Although further calculations would
certainly be needed to verify this conjecture, it is quite possible
that the observed discontinuity in the resistivity at 170~GPa
represents a structural transition from the semiconducting \aortho to
the metallic \bortho phase. Moreover, it is unclear if pseudopotential
calculations will give the accuracy needed in this transition region.

\section{Conclusions}

We have developed a new method which allows for dramatic reduction in
the number of plane waves needed in all-electron density functional
theory calculations or in calculations with hard, but highly
transferable pseudopotentials.  The foundation of the approach is to
treat the higher energy plane wave components of the electronic states
implicitly through a perturbative approach, thereby allowing these
components to be recovered quickly only when needed, thus alleviating
memory storage requirements by factors as high as fifteen in practical
calculations.  The approach lends itself well to current optimized
minimization techniques such as conjugate gradient methods and has the
benefit of allowing quantitative estimates of transferability to new
systems for which there is little or no experience and for which the
transferability of pseudopotentials is in question, such as condensed
matter at high pressure where effects in the core region become much
more important.

In addition, the first application of this approach has led to new
intriguing conjectures as to the structure of boron at high pressures.
We have carried out the first {\em ab initio} calculations of the
orthorhombic structures of boron, which we show to play an important
role in the high pressure behavior of this material.  We find that,
prior to becoming metallic, boron makes a phase transition from the
icosahedral family to the semiconducting \aortho
structure.  The low energy of this semiconducting phase indicates that
the structure of the superconducting phase for boron is more
complicated that the simple monotonic lattices explored to date.
While further calculations including optimization of the lattice
ratios for the orthorhombic phases are needed, our results lead to the
conjecture that the metallic \bortho structure is a good candidate for
the superconducting phase.

\section{Acknowledgments}
D.E.S. would like to thank Chris Henley for useful discussions, Bill
Goddard for his generous hospitality during his stay at Cal-Tech,
Peter Lepage for initiating our interest in this work and Matt Evans
for useful discussions. D.E.S. also thanks T.A.A. for support through
the MIT Department of Physics.

\appendix

\section{Application to Kleinman-Bylander
Pseudopotentials}~\label{sec-kb}

This appendix reviews briefly the application of our approach to
norm-conserving pseudopotentials of the Kleinman-Bylander form, where
the approach essentially becomes a softening procedure without
transferability biases.  The softening comes at the cost of generating
a generalized eigenvalue problem, similar in spirit to the ultrasoft
pseudopotential (USSP)~\cite{ussp} and projector augmented wave
(PAW)~\cite{paw} methods.  However, unlike the USSP and PAW
approaches, where the softening takes place referenced to a spherical
atomic environment, in our approach the softening occurs in the full
crystalline environment and thereby includes nonlocal interactions
from multiple scattering events from different atoms in the crystal
but at minimal extra computational cost.

Here, we develop the approach to work with standard separable
non-local potentials  of the form
\begin{equation}
H_{nl}(\vec G,\vec G') = \sum_{at,pr} |V_{at,pr}(\vec G) \rangle f_{pr}
\langle V_{at,pr} (\vec G')|,   \label{eqn:Vnl}
\end{equation}
where the sum is over all atoms in the cell and all projectors on a
given atom.  For such potentials it proves much more computationally
efficient to use the forms similar to
Equations~(\ref{eqn:overat}),~(\ref{eqn:CQat})~and~(\ref{eqn:Heffat})
for the overlap matrix $\mathcal{O}$, $\bfC^Q$ and $H^{eff}$,
respectively.  The overlap matrix then takes the form
\begin{eqnarray}
\mathcal{O} &=& \identity + \label{eqn:overKB} \\ &&\left( H^{PQ}_{nl}
+ H^{cr-at,PQ}_{loc}\right)\frac{1}{H^{QQ}_oH^{QQ}_o}\left(
H^{QP}_{nl} + H^{cr-at,QP}_{loc}\right), \nonumber
\end{eqnarray}
where $H^{cr-at}_{loc}$ is the local term for the ``free-atom''
crystal, a crystal whose charge density is a direct superposition of
the charge densities from isolated atoms.  This term includes the
local part of the ionic potential, the Hartree term and the
exchange-correlation term.  Ignoring the identity, expansion of
Equation~(\ref{eqn:overKB}) results in four terms, which we refer to
as nonlocal-nonlocal, nonlocal-local, local-nonlocal and local-local,
respectively.  The local-local term is exactly the same as that of the
Coulomb potential and therefore should be treated as in the
manuscript.  The advantage of employing a fixed, rather than
self-consistent, overlap matrix is that each of the remaining terms
contract contract onto the $P$-space only.  We now consider each of
these in turn.

The $\vec p,\vec p'$ component of the nonlocal-nonlocal term is
\begin{equation}
\mathcal{O}_{nl,nl}(\vec p,\vec p') = \sum_{\ind4} |V_{at,pr}(\vec p)
\rangle f_{pr} A^{(2)}_{\ind4}  f_{pr'} \langle
V_{at',pr'}(\vec p')|, \label{eqn:overnlnl}
\end{equation}
where
\begin{equation}
A^{(2)}_{\ind4} = \sum_{\vec q} \langle V_{at,pr} (\vec q)|
\left(\frac{1}{\frac{1}{2} \vec q^2}\right)^2 |V_{at',pr'}(\vec q) \rangle
\label{eqn:A2}.
\end{equation}
This term quantifies how the high-momentum states components couple to
the ions, including not only couplings to individual atoms but also to
pairs of atoms through $A^{(2)}_{\ind4}$, thereby
including environmental effects directly into the potential and
enhancing transferability.  In terms of implementation, this term
requires almost no additional memory as the only additional storage is
for $A^{(2)}$ and scales as $N_{b} \times N_{b}$, where $N_b$ is the
number of bands.  The computational demands are quite light as
$A^{(2)}$ requires an amount of computing comparable to a single
orthonormalization of the electronic bands and only need be compute
once at the very beginning of the calculation.

The local-nonlocal and nonlocal-local terms are just hermitian
conjugates of each other. The $\vec p,\vec p'$ component of the
nonlocal-local term has the form
\begin{equation}
\mathcal{O}_{nl,loc}(\vec p,\vec p') =  \sum_{at,pr} |V_{at,pr}(\vec
p) \rangle f_{pr} \langle F^{(2)}_{at,pr} (\vec p')|, \label{eqn:overlocnl}
\end{equation}
where
\begin{equation}
\langle F^{(2)}_{at,pr} (\vec p')| = \langle V_{at,pr} (\vec q)|
\left(\frac{1}{\frac{1}{2}\vec q^2}\right)^2 H^{cr-at}_{loc}(\vec q -
\vec p'). \label{eqn:F2}
\end{equation}
$F^{(2)}$ can also be simply calculated in the beginning of the
calculation through Fourier transforms of $\langle V_{at,pr}(\vec q)|
\left(\frac{1}{\frac{1}{2}\vec q^2}\right)^2|$.

The effective Hamiltonian $H^{eff}$ can also be
contracted onto the $P$-space. $H^{eff}$ has the form
$$
H^{eff} = -\left( H^{PQ}_{loc} + H^{PQ}_{nl}\right) \frac{1}{H^{QQ}_o}
\left(H^{cr-at,QP}_{loc} + H^{cr-at,QP}_{nl}\right).
$$ The nonlocal term on the left-hand side can be contracted onto
$H^{cr-at,QP}$ in a similar fashion as was done to $\mathcal{O}$,
giving two terms:
\begin{equation}
H_{nl,nl}(\vec p,\vec p') = \sum_{\ind4} |V_{at,pr}(\vec p)
\rangle f_{pr} A^{(1)}_{\ind4}  f_{pr'} \langle
V_{at',pr'}(\vec p')|, \label{eqn:Hnlnl}
\end{equation}
where
\begin{equation}
A^{(1)}_{\ind4} = \sum_{\vec q} \langle V_{at,pr} (\vec q)|
\left(\frac{1}{\frac{1}{2} \vec q^2}\right) |V_{at',pr'}(\vec q) \rangle,
\label{eqn:A1}
\end{equation}
and
\begin{equation}
H_{nl,loc}(\vec p,\vec p') =  \sum_{at,pr} |V_{at,pr}(\vec
p) \rangle f_{pr} \langle F^{(1)}_{at,pr} (\vec p')|, \label{eqn:Hlocnl}
\end{equation}
where
\begin{equation}
\langle F^{(1)}_{at,pr} (\vec p')| = \langle V_{at,pr} (\vec q)|
\left(\frac{1}{\frac{1}{2}\vec q^2}\right) H^{cr-at}_{loc}(\vec q - \vec p').
\label{eqn:F1}
\end{equation}
This will reduce almost all matrix multiplications onto the
$P$-space. The only remaining $Q$-space multiplication would be in
$H_{loc,nl}$  A final simplification that can
be made is to take
$$
H^{eff} = - H^{cr-at,PQ}\frac{1}{H^{QQ}_o}H^{cr-at,QP},
$$
and thereby make the full effective Hamiltonian contractible onto
the $P$-space.

\bibliographystyle{unsrt}

\begin{thebibliography}{10}

\bibitem{exp_rev}
For a~recent~review see:
\newblock {\em Rev. High Pressure Sci. Technol.}, 7, 1998.

\bibitem{eremets}
M.I. Eremets, V.V. Struzhkin, H.-k. Mao, and R.J. Hemley.
\newblock {\em Science}, 293:272, 2001.

\bibitem{neaton1}
J.B. Neaton and N.W. Ashcroft.
\newblock {\em Nature}, 400:141, 1999.

\bibitem{neaton2}
J.B. Neaton and N.W. Ashcroft.
\newblock {\em Phys. Rev. Lett.}, 86:2830, 2001.

\bibitem{struzhkin}
V.V. Struzhkin, R.J. Hemley, and H.K. Mao.
\newblock {\em Bull. Am. Phys. Soc.}, 44:1489, 1999.

\bibitem{fortov}
F.E. Fortov,  {\em et. al.}
\newblock {\em JETP Lett.}, 70:628, 1999.

\bibitem{hanfland}
M.~Hanfland, K.~Syassen, N.E. Christensen, and D.L. Novikov.
\newblock {\em Nature}, 408:174, 2000.

\bibitem{vohra}
Y.K. Vohra, K.E. Brister, S.~Desgreniers, A.~Ruoff, K.J. Chang, and M.L. Cohen.
\newblock {\em Phys. Rev. Lett.}, 56:1944, 1986.

\bibitem{mailhiot}
C.~Mailhiot, J.B. Grant, and A.K. McMahan.
\newblock {\em Phys. Rev. B}, 42:9033, 1990.

\bibitem{zhao}
J.~Zhao and J.P.~ Lu
\newblock {\em Phys. Rev. B}, 66:092101, 2002.

\bibitem{donohue}
J.~Donohue.
\newblock {\em The Strucutres of the Elements}, page~48.
\newblock Wiley, New York, 1974.

\bibitem{brs}
In D.~Emin, T.~Aselage, C.L. Beckel, I.A. Howard, and C.~Wood, editors, {\em
  Boron-Rich Solids, AIP Conference Proceedings 140}, Albuquerque, New Mexico,
  1986. American Institute of Physics.

\bibitem{cohen}
Ihm. J., A.~Zunger, and M.L. Cohen.
\newblock {\em J. Phys. C}, 12:4409, 1979.

\bibitem{gaussian}
B.G. Johnson, P.M.W. Gill, and J.A. Pople.
\newblock {\em J. Chem. Phys.}, 98:5612, 1993.


\bibitem{singh}
D.J. Singh.
\newblock {\em Planewaves, Pseudopotentials and the LAPW Method}.
\newblock Kluwer Academic, Norwell, MA., 1994.

\bibitem{lmto}
In H.~Dreysse, editor, {\em Electronic Structure and Physical Properties of
  Solids, The Uses of the LMTO Method}, Lectures of a Workshop Held at Mont
  Saint Odile, France, October 2-5, 1998, 2000. Springer-Verlag.

\bibitem{arias_wl}
T.A. Arias.
\newblock {\em Rev. Mod. Phys.}, 71:267, 1999.

\bibitem{barbee}
T.W. Barbee III and M.L. Cohen.
\newblock {\em Phys. Rev. B}, 44:11563, 1991.

\bibitem{natoli}
V.~Natoli, R.M. Martin, and D.M. Ceperley.
\newblock {\em Phys. Rev. Lett.}, 70:1952, 1993.

\bibitem{ballaiche}
L.~Bellaiche and K.~Kunc.
\newblock {\em Phys. Rev. B}, 55:5006, 1997.

\bibitem{hamann}
D.R. Hamann, M.~Schluter, and C.~Chiang.
\newblock {\em Phys. Rev. Lett.}, 43:1494, 1979.

\bibitem{kleinman}
L.~Kleinman and D.M. Bylander.
\newblock {\em Phys. Rev. Lett.}, 48:1425, 1982.

\bibitem{pickett}
W.~Pickett.
\newblock {\em Comput. Phys. Rep.}, 9:115, 1989.

\bibitem{teter}
M.~Teter.
\newblock {\em Phys. Rev. B}, 48:5031, 1993.

\bibitem{rappe}
I.~Grinberg, N.J. Ramer, and A.M. Rappe.
\newblock {\em Phys. Rev. B}, 63:201102, 2001.

\bibitem{louie}
S.G. Louie, S. Froyen, and M.L. Cohen.
\newblock {\em Phys. Rev, B}, 26:1738, 1982.

\bibitem{paw}
P.E. Blochl.
\newblock {\em Phys. Rev. B}, 50:17953, 1994.

\bibitem{lowdin}
P.-O. Lowdin.
\newblock {\em J. Chem. Phys.}, 19:1396, 1951.

\bibitem{feshbach}
H.~Feshbach.
\newblock {\em Ann. Phys. N.Y.}, 19:287, 1962.

\bibitem{atomic}
I.~Lindgren and J.~Morrison.
\newblock {\em Atomic Many-Body Theory}.
\newblock Springer-Verlag, Berlin, 1982.

\bibitem{joanop_ld}
D.~Vanderbilt and J.D. Joannopoulos.
\newblock {\em Phys. Rev. B}, 27:6296, 1983.

\bibitem{cohen_ld}
M.T. Yin and M.L. Cohen.
\newblock {\em Phys. Rev. B}, 26:3259, 1982.

\bibitem{cp}
R.~Car and M.~Parrinello.
\newblock {\em Phys. Rev. Lett.}, 55:2471, 1985.

\bibitem{payne}
M.C. Payne, M.P. Teter, D.C. Allen, T.A. Arias, and J.D. Joannopoulos.
\newblock {\em Rev. Mod. Phys.}, 64:1045, 1992.

\bibitem{teter_cg}
M.P. Teter, M.C. Payne, and D.C. Allan.
\newblock {\em Phys. Rev. B}, 40:12255, 1989.

\bibitem{arias_cg}
T.A. Arias, M.C. Payne, and J.D. Joannopoulos.
\newblock {\em Phys. Rev. Lett.}, 69:1077, 1992.

\bibitem{dftpp}
S.~Ismail-Beigi and T.A. Arias.
\newblock {\em Comput. Phys. Comm.}, 128:1, 2000.

\bibitem{perdew}
J.P. Perdew and A.~Zunger.
\newblock {\em Phys. Rev. B}, 23:5048, 1981.

\bibitem{gth}
S.~Goedecker, M.~Teter, and J.~Hutter.
\newblock {\em Phys. Rev. B}, 54:1703, 1996.

\bibitem{mp}
H.J. Monkhorst and J.D. Pack.
\newblock {\em Phys. Rev. B}, 13:5188, 1976.

\bibitem{ussp}
D.~Vanderbilt.
\newblock {\em Phys. Rev. B}, 41:7892, 1990.

\bibitem{fftw}
M.~Frigo and S.G. Johnson.
\newblock Fftw: An adaptive software architecture for the fft.
\newblock In {\em ICASSP conference procedings}, page 1381, 3.

\bibitem{ceperley}
D.M. Ceperley and B.J. Alder.
\newblock {\em Phys. Rev. Lett.}, 45:566, 1980.

\bibitem{zhu}
W.-J. Zhu and C.L. Henley
\newblock{\em Europhys. Lett.}, 51:133, 2000.


\bibitem{boustani1}
I. Boustani and A. Quandt and P. Kramer
\newblock{\em Europhys. Lett.}, 36:583, 1996.

\bibitem{boustani2}
I. Boustani and A. Quandt
\newblock{\em Europhys. Lett.}, 39:537, 1997.


\bibitem{wyckoff}
R.W.G. Wyckoff.
\newblock {\em Crystal Structures}, page~22-23.
\newblock 2nd ed., vol I.  Interscience Publishers, New York, c1963

\end{thebibliography}

\end{document}